\documentclass[fleqn,usenatbib,usedcolumn]{mnras}
\usepackage[british]{babel}             
\usepackage{amsmath}	
\usepackage{amssymb}	
\usepackage{mathptmx}
\let\la=\lesssim 
\usepackage[T1]{fontenc}                

\usepackage{graphicx}	

\newcommand{\division}{\mkern-\medmuskip\rotatebox[origin=c]{-15}{\scalebox{0.9}{$\Bigg/$}}\mkern-\medmuskip}

\title[Spot-induced frequency shifts]{On the contribution of sunspots to the observed frequency shifts of solar acoustic modes}

\author[A. R. G. Santos et al.]{
A. R. G. Santos$^{1,2,3}$,\thanks{E-mail: asantos@astro.up.pt}
M. S. Cunha$^{1,2}$,
P. P. Avelino$^{1,2}$,
W. J. Chaplin$^{3,4}$,
and T. L. Campante$^{3,4}$
\\
$^{1}$ Instituto de Astrof\'{i}sica e Ci\^{e}ncias do Espa\c{c}o, Universidade do Porto, CAUP, Rua das Estrelas, PT4150-762 Porto, Portugal\\
$^{2}$ Departamento de F\'{i}sica e Astronomia, Faculdade de Ci\^{e}ncias, Universidade do Porto, Rua do Campo Alegre 687, PT4169-007 Porto, Portugal\\
$^{3}$ School of Physics and Astronomy, University of Birmingham, Edgbaston, Birmingham B15 2TT, UK\\
$^{4}$ Stellar Astrophysics Centre (SAC), Department of Physics and Astronomy, Aarhus University, Ny Munkegade 120, 8000 Aarhus C, Denmark
}

\pubyear{2016}

\begin{document}
\label{firstpage}
\pagerange{\pageref{firstpage}--\pageref{lastpage}}
\maketitle

\begin{abstract}
{Activity-related variations in the solar oscillation properties have been known for 30 years. However, the relative importance of the different contributions to the observed variations is not yet fully understood.
Our goal is to estimate the relative contribution from sunspots to the observed activity-related variations in the frequencies of the acoustic modes.
We use a variational principle to relate the phase differences induced by sunspots on the acoustic waves to the corresponding changes in the frequencies of the global acoustic oscillations. From the sunspot properties (area and latitude as a function of time), we are able to estimate the spot-induced frequency shifts. These are then combined with a smooth frequency shift component, associated with long-term solar-cycle variations, and the results compared with the frequency shifts derived from the Global Oscillation Network Group (GONG) data. 
The result of this comparison is consistent with a sunspot contribution to the observed frequency shifts of roughly $30\%$, with the remaining $70\%$ resulting mostly from a global, non-stochastic variation, possibly related to the changes in the  overall magnetic field. Moreover, analysis of the residuals obtained after the subtraction of the model frequency shifts from the observations indicates the presence of a 1.5-yr periodicity in the data in phase with the quasi-biennial variations reported in the literature.}
\end{abstract}




\section{Introduction}

As a result of the solar cycle, the properties of the acoustic oscillations are observed to vary in a periodic way: as the activity level increases, the mode frequencies and the damping rates increase, while the amplitudes decrease \citep[e.g.][]{Woodard1985,Libbrecht1990a,Elsworth1990,Chaplin1998,Dziembowski2005,Metcalfe2007,Tripathy2011,Salabert2011,Salabert2015}. These activity-related variations are expected to be common in solar-like pulsators. In fact, the signature of magnetic cycles in the seismic data was detected in three solar-type stars: HD 49933 observed by the CoRoT space telescope \citep[][]{Garcia2010,Garcia2010a,Regulo2016}, and the {\it Kepler} targets KIC 10644253 \citep[][]{Salabert2016} and KIC 3733735 \citep[][]{Regulo2016}.

The activity-related variations in acoustic frequencies can result from different contributions, whose impact is still not well understood. Magnetic cycles lead to periodic changes in the global magnetic field, as well as in the total area covered by active regions, where sunspots emerge. The impact of the global magnetic field and its associated structural and thermal variations on the oscillations was addressed, e.g., by \citet{Dziembowski2005}. However, localized regions of strong magnetic field can also contribute with direct (Lorentz force) and indirect effects (thermal and structural changes).

The frequency shifts are found to be more strongly correlated with the activity proxies that are sensitive to the weak component of the magnetic field \citep[e.g.][]{Tripathy2007,Chaplin2007,Jain2009}. However, \citet{Tripathy2007} and \citet{Jain2009} argued that the strong magnetic field component also affects significantly the acoustic frequencies. This is corroborated by the frequency shifts' sensitivity to the latitudinal distribution of active regions \citep[e.g.][]{Hindman2000,Howe2002,Chaplin2004}.

The main goal of this work is to estimate the contribution of sunspots to the observed variations in the acoustic frequencies. In order to do so we follow the approach of \citet{Cunha2000} to construct a model for the frequency shifts induced by spots (Sect. \ref{sec:vprinc}). Given the sunspot properties at each epoch, the model allows for the estimation of the spot-induced frequency shifts over the solar cycle. The results obtained with real sunspot data and the corresponding comparison with the observational data are presented in Sect. \ref{sec:results}. In Sect. \ref{sec:dicussion}, we apply our model to synthetic data and we discuss the results. Section \ref{sec:conclusions} summarizes our main conclusions.

\section{Model for the spot-induced frequency shifts}\label{sec:vprinc}

The presence of sunspots on the solar photosphere can affect the propagation of acoustic waves, hence modifying the frequencies of global modes of oscillation. One way to quantify the sunspots' impact on the frequencies of global acoustic modes is by considering how the normal-mode solutions are locally perturbed by the presence of a sunspot. That perturbation can be expressed as a phase difference ($\Delta\delta$) between the solutions that would be obtained in the presence and in the absence of a spot. To fully incorporate the spot's effect, including both the direct and indirect effects, the phase must be computed at a fiducial depth, $R^*$, that is below the region of influence of the sunspot (while still sufficiently close to the surface, for the local plane-parallel approximation to be valid).  We note that ($\Delta\delta$) represents a local shift in the phase of the normal modes and should not be confused with the phase shift determined in a local-helioseismic analysis \citep[for details see][]{Cunha1998,Cunha2000}. 

Previous studies carried out in the context of strongly magnetic pulsating stars \citep[e.g.][]{Cunha2000} show that the phase difference $\Delta\delta$ depends on the magnetic field strength and inclination, as well as on the mode frequency, which affects the phase with which the wave enters the region of influence of the sunspot. However, the dependence on the mode degree is very weak, as the velocity fields of the modes of low and intermediate degree, and similar frequency, have a similar depth dependence in the very superficial layers, propagating almost vertically there, and thus interacting in a similar manner with the magnetic field.

The dependence of  $\Delta\delta$ on the properties of the magnetic field implies that this phase difference varies with the position within the spot. Nevertheless, the associated impact on the frequencies of the global oscillations is only an integrated one, which can be obtained by applying a variational principle. Based on the latter, \citet{Cunha2000} (see also \citet{Cunha1998,Cunha1999} for an application to sunspots) derived the following relation for the fractional spot-induced frequency shifts, in spherical coordinates ($r$,$\theta$,$\phi$),
\begin{equation}
\dfrac{\delta\omega}{\omega}\simeq-\dfrac{\overline{\Delta\delta}}{\omega^2\int_{r_1^{\,l}}^{R^*}c^{-2}\kappa^{-1}dr},\label{eq:fshift1}
\end{equation}
where $\omega$ is the angular frequency of the mode ($\omega=2\pi\nu$), $c$ is the sound speed, $\kappa$ is the radial component of the acoustic wave number, $r_1^{\,l}$ is the lower turning point of the mode, and $\overline{\Delta\delta}$ is the integral phase difference
\begin{equation}
\overline{\Delta\delta}=\dfrac{\int_0^{2\pi}\int_0^\pi\Delta\delta(Y_l^m)^2\sin\theta d\theta d\phi}{\int_0^{2\pi}\int_0^\pi(Y_l^m)^2\sin\theta d\theta d\phi}.\label{eq:phased}
\end{equation}
In the above, $(Y_l^m)^2=Y_l^mY_l^{m*}=(P_l^m)^2$ and $P_l^m$ are Legendre polynomials normalized such that the denominator is unity. The indices $l$ and $m$ indicate the mode angular degree and azimuthal order, respectively. At any given time, $\Delta\delta$ in Eq.~(\ref{eq:phased}) is non-zero only where spots are located.

According to Eqs.~(\ref{eq:fshift1})-(\ref{eq:phased}), the computation of the spot-induced frequency shifts from first principles would require the knowledge of the function $\Delta\delta$ within each individual sunspot. Modelling each individual sunspot over a solar cycle is clearly an impossible task, but an estimate of the spot-induced frequency shifts may still be made by considering a characteristic phase difference, $\Delta\delta_{\textrm{ch}}$, equal for all spots. This is accomplished by substituting $\Delta\delta$ in Eq.~(\ref{eq:phased}) by a function that is zero outside the sunspots and equal to a constant value $\Delta\delta_{\rm ch}$ inside the sunspots. In practice this corresponds to assuming that the sunspot area and position on the solar surface are the only properties that distinguish the impact of different sunspots on the frequencies. 

Two different approaches may be followed to estimate $\Delta\delta_{\textrm{ch}}$. The first is to consider a model for the stratification and magnetic field of a characteristic sunspot and solve the pulsation equations adequate for a plasma permeated by a strong magnetic field. 
This was performed in part in an earlier work \citep[][]{Santos2012}, where the authors considered an incomplete case in which only the indirect effect of the magnetic field on the oscillations (via the magnetically-induced changes in the stratification) was taken into account. They concluded that this indirect effect is small compared to the total spot's effect, in agreement with earlier findings by \citet{Cally2003} and \citet{Gordovskyy2007}.  While this approach may have the advantage of determining the phase difference from first principles, including the presumably dominant direct effect of the magnetic field is a rather complex task. Moreover, the results will necessarily depend on our ability to correctly model a typical sunspot. 

A second approach, which we will adopt in the current work, consists in taking $\Delta\delta_{\textrm{ch}}$ as a parameter to be constrained by direct comparison of the frequency shifts derived from Eqs.~(\ref{eq:fshift1})-(\ref{eq:phased}) and the observations. A potential difficulty in this case comes from the fact that the observed frequency shifts are not produced exclusively by the sunspots. Nevertheless, for observations with sufficient time resolution the short-term, stochastic-like frequency variations associated with the effect of the sunspots may be distinguished from the longer-term variations. This will be illustrated in Sect.~\ref{sec:results} where the relative importance of the two contributions will be established. 

To proceed, we substitute the phase difference in Eqs.~(\ref{eq:fshift1})-(\ref{eq:phased}) by the characteristic parameter $\Delta\delta_{\textrm{ch}}$. This parameter will be considered frequency-dependent, but, for the reasons explained previously, we shall neglect its dependence on mode degree. Taking the central colatitude ($\theta_i$) and longitude ($\phi_i$) of a given spot $i$, and having in mind that the spherical harmonic functions do not vary significantly within the spot, we approximate, for each spot, $\left(Y^m_l\right)_i=Y_l^m(\theta_i,\phi_i)$. Under these assumptions, Eq. (\ref{eq:fshift1}) becomes
\begin{equation}
\dfrac{\delta\omega}{\omega}\simeq-\dfrac{\Delta\delta_\textrm{ch}}{I_l}\sum_{i=1}^N\left[\left(P_l^m(\cos\theta_i)\right)^2\int_{\phi_{\textrm{min}_{\,i}}}^{\phi_{\textrm{max}_{\,i}}}\int_{\theta_{\textrm{min}_{\,i}}}^{\theta_{\textrm{max}_{\,i}}}\sin\theta d\theta d\phi\right],\label{eq:1}
\end{equation}
where $I_l=\omega^2\int_{r_1^{\,l}}^{R^*}c^{-2}\kappa^{-1}dr$ is related to the inertia of the mode and $\phi_{\textrm{min}_{\,i}}$, $\phi_{\textrm{max}_{\,i}}$, $\theta_{\textrm{min}_{\,i}}$, $\theta_{\textrm{max}_{\,i}}$ define the limits of the spot.

Considering the temporal variation of the sunspot properties, Eq. (\ref{eq:1}) can be rewritten as
\begin{equation}
\dfrac{\delta\omega}{\omega}(t)\simeq-\dfrac{\Delta\delta_\textrm{ch}}{I_lR^2}\sum_{i=1}^{N(t)}\left[\left(P_l^m(\cos\theta_i)\right)^2A_i\right],\label{eq:fshift}
\end{equation}
where $R$ is the solar radius and $A_i$ the area of a given spot $i$. In addition to the temporal variation, the fractional frequency shifts given by Eq.~(\ref{eq:fshift}) depend on mode frequency, due to the frequency dependence of $\Delta\delta_{\textrm{ch}}$, and on mode degree, due to the degree dependence of the Legendre polynomial and the degree dependence of the mode inertia (reflected in the integral $I_{\it l}$).

\section{Results}\label{sec:results}

In order to study the contribution of the sunspots to the observed activity-related frequency shifts, we have used the sunspot daily records from the National Geophysical Data Center (NGDC/NOAA). This database includes daily information about each observed sunspot group, such as latitude and area. With the sunspot group area and latitude in hand, Eq.~(\ref{eq:fshift}) allows for the estimation of the spot-induced frequency shifts for different radial orders, angular degrees, and azimuthal orders, for any given characteristic phase difference.

In this work the model frequency shifts will be compared with the observational data from GONG presented in \citet{Tripathy2011}. In the latter, the acoustic frequencies were obtained with a cadence of 36 days and an overlap of 18 days. \citet{Tripathy2011} used in their calculations the modes of degree between $l=0$ and $100$ with frequencies ranging from $2000$ to $3300\,\mu$Hz that are present in all the sub-datasets. For each multiplet the authors first derived a central frequency. The central frequency shifts, defined by comparison with a reference value, were then combined, weighted by the corresponding mode inertias, to establish a mean frequency shift for the observed frequency range.

Within the frequency range considered by \citet{Tripathy2011}, the observed frequency shifts are almost a linear function of the frequency \citep[e.g][]{Libbrecht1990a,Chaplin2001,Basu2002}. Therefore, the mean frequency shifts derived by the authors provides a good estimate of the frequency shifts of modes with frequencies close to the middle of the observed interval. With this in mind, in the computation of the model frequency shifts we will consider the frequency closest to the center of the frequency range used by \citet{Tripathy2011} \citep[frequencies from][]{Rabello-Soares1999}. This implies that the phase difference $\Delta\delta_{\rm ch}$ that will be inferred from comparison of our model with the observations must be interpreted as the characteristic phase difference at that frequency. 

The NGDC/NOAA data includes only the sunspots emerging on the visible side of the Sun. Since all sunspots on the solar surface, and not only the visible ones, contribute to the frequency shifts, we must make an assumption about the contribution of the invisible spots. As there is no reason to expect systematic differences between the two sides, it is reasonable to assume that on average the contributions are equal. Therefore, the fractional frequency shifts will be computed from Eq. (\ref{eq:fshift}) by taking twice the value obtained when only the visible spots are considered. The impact of doing so will be discussed at length in Sect. \ref{sec:dicussion}, where we will repeat our analysis using synthetic solar-cycle data. The spot-induced frequency shifts, $\delta\nu_{lm}=\delta\omega_{lm}/2\pi$, are computed for $l=0-100$ and the corresponding azimuthal orders. In this calculation we have taken $R^*$ to correspond to a radius of $10$ Mm below the photosphere.
The model frequency shift associated to a given angular degree, $\delta\nu_l$, is then calculated following the same procedure used by \citet{Tripathy2007,Tripathy2011}. In particular, we fit the $\delta\nu_{lm}$ to a polynomial expansion of the form
\begin{equation}
\delta\nu_{lm}=\delta\nu_l+\sum_{j=1}^{j_\textrm{max}}a_j(l)P_j(m/l),
\label{eq:fref}\end{equation}
where $j_\textrm{max}$ is $2l$ for $l\leq4$ and $9$ for $l>4$, $a_j$ are the splitting coefficients  and $P_j$ is the Legendre polynomial of order $j$.

The model mean frequency shifts $\delta\nu(t)$ are then defined as
\begin{equation}
\delta\nu(t)=\sum_lQ_l\delta\nu_l(t)\division\sum_lQ_l\equiv\delta\nu_\textrm{spots},
\end{equation}
where $Q_l$ is the inertia ratio \citep[$E_l/E_0(\nu_l)$;][]{Christensen-Dalsgaard1991}. $\delta\nu_\textrm{spots}$ is thus the
model equivalent to the observable constructed by \citet{Tripathy2011} based on the GONG data, but including only the spot-induced part of the frequency shifts. For a known sunspot distribution, it depends on the single parameter $\Delta\delta_\textrm{ch}$.

\begin{figure}
\includegraphics[width=\hsize]{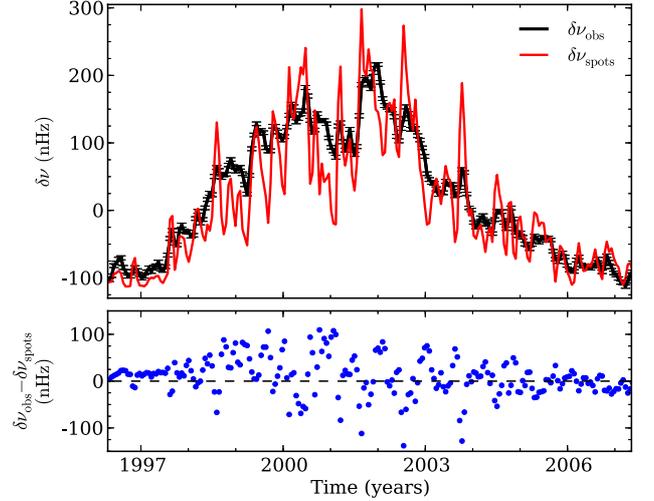}
\caption{Top panel: Observed \citep[black;][]{Tripathy2011} and model spot-induced frequency shifts (red) for the hypothetical case in which spots are taken to be the only contribution for the observed frequency shifts (see text for details). The amplitude of the variations in $\delta\nu$ is much larger in our results than in the observational data, indicating that the spots are not the only contribution to the frequency shifts. Bottom panel: Residuals between the observed and model spot-induced frequency shifts.}\label{fig:fspots}
\end{figure}

It is well known that the observed frequency shifts do not result only from the effect of the sunspots \citep[e.g.][]{Tripathy2007,Chaplin2007,Jain2009}. Other effects, such as the variation of the global magnetic field and structural and thermal changes act on much longer time-scales. As a consequence, in order to reproduce the observations it is necessary to add to our model an additional component of the frequency shifts, varying on longer time-scales. That will be done later in this section. Nevertheless, it is instructive to consider, for a moment, the hypothetical case in which spots are the sole responsible source of the observed frequency shifts. In this case, the characteristic phase difference may be determined by fitting the model for the spot-induced frequency shifts to the observational data ($\delta\nu_\textrm{obs}$). To that end, we obtain the 36-day averages (with an overlap of 18 days) of the daily spot-induced frequency shifts. Then, using the consecutive independent data points\footnote{Since the observations have a cadence of 36 days with an overlap of 18 days, consecutive data points are not independent. In the $\chi^2$ minimization, we have thus used only every second data point and the corresponding results from the model.}, the best fit is obtained through a $\chi^2$ minimization between the observational values and the model values, $\delta\nu_\textrm{spots}$. We find that $\Delta\delta_\textrm{ch}\!\sim\!-1.51$ and the corresponding model-data comparison is shown in Fig. \ref{fig:fspots}. 

As seen from Fig. \ref{fig:fspots}, the amplitude of the frequency shifts' variations is about three times larger in the model than in the observed data. This is a consequence of having taken the spots to be the sole responsible for the observed frequency shifts, which, as mentioned earlier, is well known not to be the case. In other words, while sunspots contribute both to the long and short-term variations in the oscillation frequencies, if they were the single cause for the frequency shifts, the observations would show a much more significant short-term variance.  We can thus ask the question of how significant would a long-term-varying component originating from sources other than the sunspots have to be, in order to adequately fit the observations. To address this question, we add the long-term-varying component ($\delta\nu_\textrm{global}$) to the frequency shifts, which may be associated to the effect of the overall solar magnetic field and to global structural and thermal changes. Different approaches may be used to define the smooth component. Given that the long-term frequency shifts variations are well correlated with the long-term variations in the sunspot number \citep[e.g.][]{Tripathy2007,Chaplin2007,Jain2009,Jain2012}, one option is to use the function that was proposed by \citet{Hathaway1994} to fit the sunspot number. Accordingly, we fit the observed frequency shifts with the function
\begin{equation}
f(t)=\dfrac{A(t-t_0)^3}{\exp((t-t_0)^2/B^2)-C},
\end{equation}
where $A$ is the amplitude, $t_0$ is the starting time, $B$ is related to the size of the rising phase and $C$ is the asymmetry parameter. The values of the parameters found from the fit are: $A\sim6.76$, $B\sim4.71$, $C\sim-0.46$, and $t_0\sim1995.06$.

The model frequency shifts are then given by
\begin{equation}
\delta\nu_\textrm{model}=\delta\nu_\textrm{global}+\delta\nu_\textrm{spots},\label{eq:comb2}
\end{equation}
where $\delta\nu_\textrm{global}=wf(t)$ and $w$ is the weight of the smooth component, to be defined by fitting the data. Thus, our model contains two parameters, $w$ and $\Delta\delta_\textrm{ch}$.

From a $\chi^2$ minimization between $\delta\nu_{\rm model}$ and the observations, we find $w\sim0.71$ and $\Delta\delta_{\rm ch}\sim-0.44$. The latter can be used to estimate the travel-time perturbation induced by a sunspot (relative to the quiet sun). An explicit relation between the double-skip travel-time perturbation, $\tau_2$ \citep[e.g.][]{Zhao2006} and the phase shift of normal modes has been derived by \citet[eq. 8]{Cunha1998}. Using the value of $\Delta\delta_{\rm ch}$ and the frequency of the center of the observed interval in this equation, we estimate the one-half of the double-skip travel-time perturbation to be $\tau_2/2\sim-0.37\, \rm min$.

Figure 2 shows the comparison between the observed frequency shifts and the results for $\delta\nu_{\rm model}$ with the parameters derived above. The resulting frequency shifts are in reasonable agreement with the observational data indicating that our simple, two-parameter model captures the main features contained in the observed frequency shifts. The weight of the smooth component, $w$, gives an estimate of the contribution resulting from the changes in the global magnetic field and associated structural and thermal variations, which is $\sim70\%$. The remaining $\sim30\%$ correspond then to the spot-induced contribution.

\begin{figure}
\centering
\includegraphics[width=\hsize]{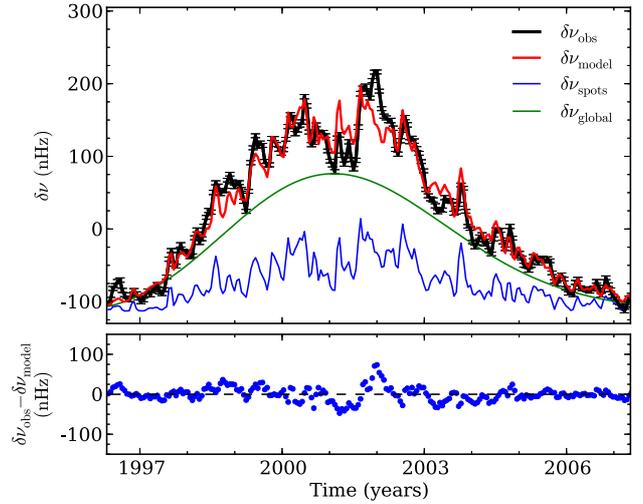}
\caption{Top panel: Observed \citep[black;][]{Tripathy2011} and model frequency shifts (red). The total model frequency shifts correspond to the combination of a global (green) and a spot-induced (blue) components. Bottom panel: Residuals between the observed and model frequency shifts.}\label{fig:tp1}
\end{figure}

\section{Discussion}\label{sec:dicussion}

Different approaches may be considered to model the smooth component introduced in Sect.~\ref{sec:results}. While we have opted to use the function defined by \citet{Hathaway1994}, we have also assessed the robustness of our results to changes in the modelling of the smooth component. We considered an alternative approach in which the function {\it f(t)} is defined by a smoothed version of the observed frequency shifts obtained through the application of a 360-day filter.  Moreover, we have also considered an alternative approach to the model-data comparison, by which we have first subtracted a smooth component to both the observed and the spot-induced frequency shifts (obtained through the application of a 360-day filter to each set) and, then, fitted the residuals. In both cases the global and spot-induced fractional contributions were found to be similar to the values estimated in the original analysis, with the parameter $w$ agreeing to better than 10$\%$ with the value estimated in Sect.~\ref{sec:results}.

While the results from our two-parameter model reproduce reasonably well the observed frequency shifts, we see from the lower panel of Fig. \ref{fig:tp1} that some differences still exist between the two. Part of these residuals may be connected to the fact that only the visible sunspot groups are recorded in the NGDC/NOAA daily records.
The observed frequency shifts are affected by the spots appearing throughout the whole solar surface.  Thus, using the visible groups in the computation of the spot-induced frequency shifts from Eq. (\ref{eq:fshift}) and considering that the total spot-induced component is twice the value may introduce statistical differences between the model predictions and the observations. Since there is no data for the sunspots emerging on the invisible side of the Sun, the only way to check the impact of this limitation is to recur to simulations, which we will do by performing the same analysis as in Sect. \ref{sec:results} but using synthetic sunspot records obtained with the empirical tool developed by \citet{Santos2015}.  With this tool we simulate the number of sunspot groups, the total area covered by them and their latitudinal distribution along one complete solar cycle. The output from this tool is analogous to that of the real daily records.

\begin{figure}
\includegraphics[width=\hsize]{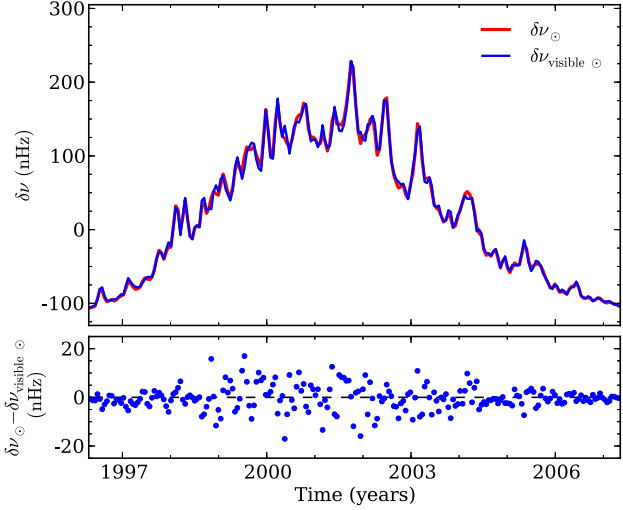}
\caption{Top panel: Frequency shifts derived from synthetic sunspot data: when considering all synthetic sunspot groups at the solar surface (red) and when considering only the visible groups (blue). Bottom panel: Residuals between the frequency shifts shown in the top panel.}\label{fig:simul}
\end{figure}

Following the methodology described in Sect. \ref{sec:results}, we obtain the daily frequency shifts induced by all the groups on the solar surface (case 1) and by the visible groups alone assuming that they are a reasonable representation of both (visible and invisible) sides of the Sun (case 2). The first of these cases is used to simulate a model frequency-shift cycle, i.e. a model equivalent of the observations considered in the previous section, while the second is used to estimate the error made when taking only the visible sunspot groups for the frequency shift calculations, with a factor of two to account for the invisible side. Since we want our model-simulated cycle to be as similar to the real data as possible, in the first case the parameter $\Delta\delta_\textrm{ch}$ needed to compute $\delta\nu_{\rm spots}$ is determined in such a way that the sum of the variations of the frequency shifts obtained from synthetic data coincides with the value found using the observed frequency shifts, i.e. $\sum_k|\Delta\delta\nu|_{\textrm{obs},k}\!=\!\sum_k|\Delta\delta\nu|_{\textrm{synt-spots},k}$, where $\Delta\delta\nu=\delta\nu_{k+1}-\delta\nu_{k}$ and different indices $k$ correspond to the different data points. To the synthetic $\delta\nu_{\rm spots}$ we add the smooth component found in Sect. \ref{sec:results}, with a weight $w$, to obtain the model frequency shifts for case 1 (hereafter $\delta\nu_{\odot}$). The weight $w$ is determined through a $\chi^2$ minimization with the observed data. The frequency shifts $\delta\nu_\odot$ are now taken as the reference with which we compare the results obtained from the visible spots alone. We then proceed exactly as in Sect. \ref{sec:results}, but substituting the observed frequency shifts by the reference-model frequency shifts, and the frequency shifts based on sunspot data from NGDC/NOAA by the frequency shifts obtained from simulated data for the visible side alone, hereafter $\delta\nu_{\rm visible\,\odot}$. 
Figure \ref{fig:simul} shows the comparison between $\delta\nu_{\odot}$ and $\delta\nu_{\rm visible\,\odot}$ for a given simulation of the sunspot cycle. 

The differences between the model frequency shifts derived when considering all synthetic sunspot groups at the surface and when considering just the visible groups (in Fig. \ref{fig:simul}) are significantly smaller and more evenly distributed than the ones found in Sect. \ref{sec:results}. We, therefore, conclude that the main differences seen in Fig. \ref{fig:tp1} are not explained by the fact that the sunspot data used for the computation of the model frequency shifts in Sect. \ref{sec:results} only contain information on the groups emerging on the visible side of the Sun.

Since the simulations of the sunspot cycles are stochastic, we repeated the procedure described above for a large number of synthetic data sets. We found that the spot-induced contributions to the frequency shifts obtained when considering all synthetic sunspot groups and the visible groups alone, differ, on average, by less than $3\%$. The difference is small due to the 36-day averaging of the frequency shifts. This average dilutes the daily differences that may exist between the area and position of the spots in each side of the Sun. This reassures us that the estimated value for the spot-induced frequency shifts' contribution found in Sect. \ref{sec:results} is robust, even if computed from data for the visible sunspots only.

\begin{figure}
\includegraphics[width=\hsize]{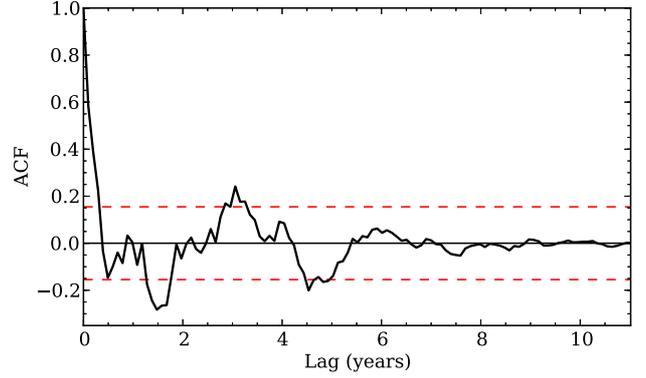}
\caption{Autocorrelation function of the residuals in Fig. \ref{fig:tp1}. The red dashed lines indicate the $95\%$ significance level.}\label{fig:acf}
\end{figure}

A closer inspection  of the residuals shows that these seem to be in phase with the quasi-biennial variations found in the frequency shifts by \citet{Fletcher2010} and \citet{Broomhall2012}, having a significant periodicity of $\sim\!\!1.5$ years (Fig. \ref{fig:acf}). A more detailed analysis indicates that the 1.5-yr periodicity is intrinsic to the observed frequency shifts.

This may point to a global contribution with structure in an intermediate time-scale, as opposed to one that varies only on the time-scale of the solar cycle, as considered in Sect.~\ref{sec:results}. In addition, the characteristic phase difference may vary during the solar cycle, e.g. as a result of changes in the typical structure and magnetic field within the sunspots. This could also contribute to the differences seen in Fig. \ref{fig:tp1}.

\section{Conclusions}\label{sec:conclusions}

Our results indicate that there are two main contributions to the total activity-related frequency shifts. A spot-induced contribution of about $30\%$,  that is responsible for the stochastic behaviour of the frequency shifts as well as for part of their long-term variations, and a global contribution of about $\sim\!\!70\%$, varying on the time-scale of the solar cycle, possibly related to the changes in the overall magnetic field. This is consistent with the results obtained by \citet{Jain2009} when studying the correlations between the frequency shifts and different activity indices. Furthermore, the characteristic phase shift derived from the fit of our two-parameter model to the data translates into a travel-time perturbation that is consistent with those inferred from time-distance helioseismology \citep[][]{Duvall1995,Duvall1996,Zhao2006}.

We have checked that our results are robust against changes in the form adopted for the global component. Moreover, based on synthetic data, we have demonstrated that they would not change in any significant way, if all the sunspots emerging on the surface of the Sun could be considered, rather than just the visible ones.

Although, our model frequency shifts agree reasonably well with the observed frequency shifts, we find evidence for variations in the data at time-scales not accounted for in our model. These could either be associated with intermediate-term variations in the spot-induced phase shifts or with intermediate-term variations in the non-spot contributions. Despite these differences, we note that the contributions of $70\%$ and $30\%$ derived from our model are robust, as they depend essentially on the relative amplitudes of the stochastic and smooth components seen in the observations. In fact, we also verified by testing models with an additional intermediate-term component, that small corrections to the model associated to activity on intermediate time-scales will not change the estimated fractional contribution of the spot-induced and global components in any significant way.

With the increasing number of solar-like stars with detected activity-related variations in the frequencies of the acoustic modes \citep[][]{Garcia2010,Salabert2016,Regulo2016}, the understanding of the different contributions to those variations becomes even more important. If the contribution from starspots to the frequency shifts in solar-like pulsators is comparable to that found in this work for the case of the Sun, the amplitude of those shifts can be used to estimate the surface coverage of starspots.

\section*{Acknowledgements}

This work was supported by Funda\c{c}\~{a}o para a Ci\^{e}ncia e a Tecnologia (FCT) through the research grant UID/FIS/04434/2013. ARGS acknowledges the support from FCT through the Fellowship SFRH/BD/88032/2012 and from the University of Birmingham. MSC and PPA acknowledge support from FCT through the Investigador FCT Contracts No. IF/00894/2012 and IF/00863/2012 and POPH/FSE (EC) by FEDER funding through the program Programa Operacional de Factores de Competitividade (COMPETE). TLC and WJC acknowledge the support of the UK Science and Technology Facilities Council (STFC). The research leading to these results has received funding from EC, under FP7, through the grant agreement FP7-SPACE-2012-312844 and PIRSES-GA-2010-269194. ARGS, MSC, and PPA are grateful for the support from the High Altitude Observatory (NCAR/UCAR), where part of the current work was developed.




\bibliographystyle{mnras}
\bibliography{fshifts} 

\begin{thebibliography}{}
\makeatletter
\relax
\def\mn@urlcharsother{\let\do\@makeother \do\$\do\&\do\#\do\^\do\_\do\%\do\~}
\def\mn@doi{\begingroup\mn@urlcharsother \@ifnextchar [ {\mn@doi@}
  {\mn@doi@[]}}
\def\mn@doi@[#1]#2{\def\@tempa{#1}\ifx\@tempa\@empty \href
  {http://dx.doi.org/#2} {doi:#2}\else \href {http://dx.doi.org/#2} {#1}\fi
  \endgroup}
\def\mn@eprint#1#2{\mn@eprint@#1:#2::\@nil}
\def\mn@eprint@arXiv#1{\href {http://arxiv.org/abs/#1} {{\tt arXiv:#1}}}
\def\mn@eprint@dblp#1{\href {http://dblp.uni-trier.de/rec/bibtex/#1.xml}
  {dblp:#1}}
\def\mn@eprint@#1:#2:#3:#4\@nil{\def\@tempa {#1}\def\@tempb {#2}\def\@tempc
  {#3}\ifx \@tempc \@empty \let \@tempc \@tempb \let \@tempb \@tempa \fi \ifx
  \@tempb \@empty \def\@tempb {arXiv}\fi \@ifundefined
  {mn@eprint@\@tempb}{\@tempb:\@tempc}{\expandafter \expandafter \csname
  mn@eprint@\@tempb\endcsname \expandafter{\@tempc}}}

\bibitem[\protect\citeauthoryear{Basu}{Basu}{2002}]{Basu2002}
Basu S.,  2002. eprint: arXiv:astro-ph/0205049, pp 7--14

\bibitem[\protect\citeauthoryear{Broomhall, Chaplin, Elsworth  \&
  Simoniello}{Broomhall et~al.}{2012}]{Broomhall2012}
Broomhall A.-M.,  Chaplin W.~J.,  Elsworth Y.,   Simoniello R.,  2012, \mn@doi
  [MNRAS] {10.1111/j.1365-2966.2011.20123.x}, 420, 1405

\bibitem[\protect\citeauthoryear{Cally, Crouch  \& Braun}{Cally
  et~al.}{2003}]{Cally2003}
Cally P.~S.,  Crouch A.~D.,   Braun D.~C.,  2003, \mn@doi [MNRAS]
  {10.1046/j.1365-2966.2003.07019.x}, 346, 381

\bibitem[\protect\citeauthoryear{Chaplin, Elsworth, Isaak, Lines, McLeod,
  Miller  \& New}{Chaplin et~al.}{1998}]{Chaplin1998}
Chaplin W.~J.,  Elsworth Y.,  Isaak G.~R.,  Lines R.,  McLeod C.~P.,  Miller
  B.~A.,   New R.,  1998, \mn@doi [MNRAS] {10.1046/j.1365-8711.1998.01999.x},
  300, 1077

\bibitem[\protect\citeauthoryear{Chaplin, Appourchaux, Elsworth, Isaak  \&
  New}{Chaplin et~al.}{2001}]{Chaplin2001}
Chaplin W.~J.,  Appourchaux T.,  Elsworth Y.,  Isaak G.~R.,   New R.,  2001,
  \mn@doi [MNRAS] {10.1046/j.1365-8711.2001.04357.x}, 324, 910

\bibitem[\protect\citeauthoryear{Chaplin, Elsworth, Isaak, Miller  \&
  New}{Chaplin et~al.}{2004}]{Chaplin2004}
Chaplin W.~J.,  Elsworth Y.,  Isaak G.~R.,  Miller B.~A.,   New R.,  2004,
  \mn@doi [MNRAS] {10.1111/j.1365-2966.2004.07998.x}, 352, 1102

\bibitem[\protect\citeauthoryear{Chaplin, Elsworth, Miller, Verner  \&
  New}{Chaplin et~al.}{2007}]{Chaplin2007}
Chaplin W.~J.,  Elsworth Y.,  Miller B.~A.,  Verner G.~A.,   New R.,  2007,
  \mn@doi [ApJ] {10.1086/512543}, 659, 1749

\bibitem[\protect\citeauthoryear{Christensen-Dalsgaard \&
  Berthomieu}{Christensen-Dalsgaard \&
  Berthomieu}{1991}]{Christensen-Dalsgaard1991}
Christensen-Dalsgaard J.,  Berthomieu G.,  1991, in , Solar {{Interior}} and
  {{Atmosphere}}.
pp 401--478

\bibitem[\protect\citeauthoryear{{Cunha}}{{Cunha}}{1999}]{Cunha1999}
{Cunha} M.~S.,  1999, PhD thesis, Cambrigde University, UK

\bibitem[\protect\citeauthoryear{Cunha \& Gough}{Cunha \&
  Gough}{2000}]{Cunha2000}
Cunha M.~S.,  Gough D.,  2000, \mn@doi [MNRAS]
  {10.1046/j.1365-8711.2000.03896.x}, 319, 1020

\bibitem[\protect\citeauthoryear{Cunha, Br{\"u}ggen  \& Gough}{Cunha
  et~al.}{1998}]{Cunha1998}
Cunha M.~S.,  Br{\"u}ggen M.,   Gough D.~O.,  1998. eprint:
  arXiv:astro-ph/9807123, p.~905

\bibitem[\protect\citeauthoryear{Duvall}{Duvall}{1995}]{Duvall1995}
Duvall Jr. T.~L.,  1995. {ASP Conf. Ser. 76}, p.~465

\bibitem[\protect\citeauthoryear{Duvall, D'Silva, Jefferies, Harvey  \&
  Schou}{Duvall et~al.}{1996}]{Duvall1996}
Duvall Jr. T.~L.,  D'Silva S.,  Jefferies S.~M.,  Harvey J.~W.,   Schou J.,
  1996, \mn@doi [Nature] {10.1038/379235a0}, 379, 235

\bibitem[\protect\citeauthoryear{Dziembowski \& Goode}{Dziembowski \&
  Goode}{2005}]{Dziembowski2005}
Dziembowski W.~A.,  Goode P.~R.,  2005, \mn@doi [ApJ] {10.1086/429712}, 625,
  548

\bibitem[\protect\citeauthoryear{Elsworth, Howe, Isaak, McLeod  \&
  New}{Elsworth et~al.}{1990}]{Elsworth1990}
Elsworth Y.,  Howe R.,  Isaak G.~R.,  McLeod C.~P.,   New R.,  1990, \mn@doi
  [Nature] {10.1038/345322a0}, 345, 322

\bibitem[\protect\citeauthoryear{Fletcher, Broomhall, Salabert, Basu, Chaplin,
  Elsworth, Garcia  \& New}{Fletcher et~al.}{2010}]{Fletcher2010}
Fletcher S.~T.,  Broomhall A.-M.,  Salabert D.,  Basu S.,  Chaplin W.~J.,
  Elsworth Y.,  Garcia R.~A.,   New R.,  2010, \mn@doi [ApJ]
  {10.1088/2041-8205/718/1/L19}, 718, L19

\bibitem[\protect\citeauthoryear{Garc{\'\i}a, Mathur, Salabert, Ballot, Regulo,
  Metcalfe  \& Baglin}{Garc{\'\i}a et~al.}{2010a}]{Garcia2010}
Garc{\'\i}a R.~A.,  Mathur S.,  Salabert D.,  Ballot J.,  Regulo C.,  Metcalfe
  T.~S.,   Baglin A.,  2010a, \mn@doi [Science] {10.1126/science.1191064}, 329,
  1032

\bibitem[\protect\citeauthoryear{Garc{\'\i}a, Ballot, Mathur, Salabert  \&
  Regulo}{Garc{\'\i}a et~al.}{2010b}]{Garcia2010a}
Garc{\'\i}a R.~A.,  Ballot J.,  Mathur S.,  Salabert D.,   Regulo C.,  2010b,
  preprint, 1012, arXiv:1012.0494

\bibitem[\protect\citeauthoryear{Gordovskyy \& Jain}{Gordovskyy \&
  Jain}{2007}]{Gordovskyy2007}
Gordovskyy M.,  Jain R.,  2007, \mn@doi [ApJ] {10.1086/513737}, 661, 586

\bibitem[\protect\citeauthoryear{Hathaway, Wilson  \& Reichmann}{Hathaway
  et~al.}{1994}]{Hathaway1994}
Hathaway D.~H.,  Wilson R.~M.,   Reichmann E.~J.,  1994, \mn@doi [Sol. Phys.]
  {10.1007/BF00654090}, 151, 177

\bibitem[\protect\citeauthoryear{Hindman, Haber, Toomre  \& Bogart}{Hindman
  et~al.}{2000}]{Hindman2000}
Hindman B.,  Haber D.,  Toomre J.,   Bogart R.,  2000, \mn@doi [Sol. Phys.]
  {10.1023/A:1005283302728}, 192, 363

\bibitem[\protect\citeauthoryear{Howe, Komm  \& Hill}{Howe
  et~al.}{2002}]{Howe2002}
Howe R.,  Komm R.~W.,   Hill F.,  2002, \mn@doi [ApJ] {10.1086/343892}, 580,
  1172

\bibitem[\protect\citeauthoryear{Jain, Tripathy  \& Hill}{Jain
  et~al.}{2009}]{Jain2009}
Jain K.,  Tripathy S.~C.,   Hill F.,  2009, \mn@doi [ApJ]
  {10.1088/0004-637X/695/2/1567}, 695, 1567

\bibitem[\protect\citeauthoryear{Jain, Tripathy, Watson, Fletcher, Jain  \&
  Hill}{Jain et~al.}{2012}]{Jain2012}
Jain R.,  Tripathy S.~C.,  Watson F.~T.,  Fletcher L.,  Jain K.,   Hill F.,
  2012, \mn@doi [A\&A] {10.1051/0004-6361/201219876e}, 545, A73

\bibitem[\protect\citeauthoryear{Libbrecht \& Woodard}{Libbrecht \&
  Woodard}{1990}]{Libbrecht1990a}
Libbrecht K.~G.,  Woodard M.~F.,  1990, \mn@doi [Nature] {10.1038/345779a0},
  345, 779

\bibitem[\protect\citeauthoryear{Metcalfe, Dziembowski, Judge  \&
  Snow}{Metcalfe et~al.}{2007}]{Metcalfe2007}
Metcalfe T.~S.,  Dziembowski W.~A.,  Judge P.~G.,   Snow M.,  2007, \mn@doi
  [MNRAS] {10.1111/j.1745-3933.2007.00325.x}, 379, L16

\bibitem[\protect\citeauthoryear{Rabello-Soares \& Appourchaux}{Rabello-Soares
  \& Appourchaux}{1999}]{Rabello-Soares1999}
Rabello-Soares M.~C.,  Appourchaux T.,  1999, A\&A, 345, 1027

\bibitem[\protect\citeauthoryear{R{\'e}gulo, Garc{\'\i}a  \& Ballot}{R{\'e}gulo
  et~al.}{2016}]{Regulo2016}
R{\'e}gulo C.,  Garc{\'\i}a R.~A.,   Ballot J.,  2016, preprint, 1603,
  arXiv:1603.04673

\bibitem[\protect\citeauthoryear{Salabert, Garc{\'\i}a, Pall{\'e}  \&
  Jim{\'e}nez}{Salabert et~al.}{2011}]{Salabert2011}
Salabert D.,  Garc{\'\i}a R.~A.,  Pall{\'e} P.~L.,   Jim{\'e}nez A.,  2011,
  \mn@doi [J. Phys. Conf. Ser.] {10.1088/1742-6596/271/1/012030}, 271, 012030

\bibitem[\protect\citeauthoryear{Salabert, Garc{\'\i}a  \&
  Turck-Chi{\`e}ze}{Salabert et~al.}{2015}]{Salabert2015}
Salabert D.,  Garc{\'\i}a R.~A.,   Turck-Chi{\`e}ze S.,  2015, \mn@doi [A\&A]
  {10.1051/0004-6361/201425236}, 578, A137

\bibitem[\protect\citeauthoryear{Salabert et~al.,}{Salabert
  et~al.}{2016}]{Salabert2016}
Salabert D.,  et~al., 2016, preprint, 1603, arXiv:1603.00655

\bibitem[\protect\citeauthoryear{Santos, Cunha  \& Lima}{Santos
  et~al.}{2012}]{Santos2012}
Santos A. R.~G.,  Cunha M.~S.,   Lima J. J.~G.,  2012, \mn@doi [Astron. Nachr.]
  {10.1002/asna.201211821}, 333, 1032

\bibitem[\protect\citeauthoryear{Santos, Cunha, Avelino  \& Campante}{Santos
  et~al.}{2015}]{Santos2015}
Santos A. R.~G.,  Cunha M.~S.,  Avelino P.~P.,   Campante T.~L.,  2015, \mn@doi
  [A\&A] {10.1051/0004-6361/201425299}, 580, A62

\bibitem[\protect\citeauthoryear{Tripathy, Hill, Jain  \& Leibacher}{Tripathy
  et~al.}{2007}]{Tripathy2007}
Tripathy S.~C.,  Hill F.,  Jain K.,   Leibacher J.~W.,  2007, \mn@doi [Sol.
  Phys.] {10.1007/s11207-007-9000-z}, 243, 105

\bibitem[\protect\citeauthoryear{Tripathy, Jain, Salabert, Garc{\'\i}a, Hill
  \& Leibacher}{Tripathy et~al.}{2011}]{Tripathy2011}
Tripathy S.~C.,  Jain K.,  Salabert D.,  Garc{\'\i}a R.~A.,  Hill F.,
  Leibacher J.~W.,  2011, \mn@doi [J. Phys. Conf. Ser.]
  {10.1088/1742-6596/271/1/012055}, 271, 012055

\bibitem[\protect\citeauthoryear{Woodard \& Noyes}{Woodard \&
  Noyes}{1985}]{Woodard1985}
Woodard M.~F.,  Noyes R.~W.,  1985, \mn@doi [Nature] {10.1038/318449a0}, 318,
  449

\bibitem[\protect\citeauthoryear{Zhao \& Kosovichev}{Zhao \&
  Kosovichev}{2006}]{Zhao2006}
Zhao J.,  Kosovichev A.~G.,  2006, \mn@doi [ApJ] {10.1086/503248}, 643, 1317

\makeatother
\end{thebibliography}

\bsp	
\label{lastpage}
\end{document}